\documentclass[]{article}
\usepackage[latin1]{inputenc}
\usepackage{latexsym}
\usepackage{graphicx}
\usepackage[pdftex=true,colorlinks=true,plainpages=false]{}

\title{{\bf Note on Hawking-Unruh effects in Graphene}}
\author{Pisin Chen and Haret Rosu\footnote{Permanent address: IPICyT, Instituto Potosino de Investigacion Cientifica y Tecnologica, Apdo Postal 3-74 Tangamanga, 78231 San Luis Potos\'{\i}, Mexico}\\
{\em Leung Center for Cosmology and Particle Astrophysics and}\\
{\em Department of Physics and Graduate Institute of Astrophysics}\\
{\em National Taiwan University, Taipei, 10617 Taiwan}\\
{\small  Emails: pisinchen@phys.ntu.edu.tw and hcr@ipicyt.edu.mx}}
\date{{\tt Mod. Phys. Lett. A 27(37) (2012) 1250218\\
{\small doi: 10.1142/S0217732312502185}}}

\begin{document}
\maketitle

\begin{abstract}
\noindent Beltrami-shaped graphene sheets have been recently proposed as analogs of curved spacetimes with Hawking-Unruh effects detected through  typical condensed matter measurements involving scanning tunneling microscopes and spectroscopy. However, such deformed sheets, if ever fabricated, will contain large strain-induced pseudomagnetic fields with important guiding effects on the motion of the electrons in the conduction band. Besides, possible surface polariton and plasmon modes are known to be important players in the radiative heat transfer which takes place in the natural near-field nanoscale experimental conditions. Therefore, we suggest that the latter class of experiments could shed light on phenomena related to the black hole membrane paradigm instead. \\


\end{abstract}

Hawking thermal radiation from black hole event horizons and the similar Unruh effect in the case of accelerated elementary particles have been stimulating scientific paradigms for the last four decades as two of the most exquisite results in theoretical physics. However, experimental evidence of these fundamental effects is so far very poor if not lacking because they are extremely weak in standard terrestrial and astrophysical conditions. Nevertheless, in the early 1980s, the first proposals for experimental detection of analog thermal effects in terrestrial laboratories, including accelerators, came out one after the other, starting with Unruh's acoustic thermal bursts in supersonic hydrodynamic flows.
This research area has developed steadily, especially during the last decade when a plethora of promising setups have been devised. Now, the analogue (curved) spacetime program is vigorous and exciting for the development of future theory and technology. A large amount of information regarding this program has been accumulated in the living review of Visser and collaborators initiated in 2005 and updated in 2011 \cite{livrev}.

\bigskip

A very recent analogue spacetime, introduced by Iorio and Lambiase \cite{il-2011}, is hosted by graphene, the 2010 Nobel Prize nanomaterial. These authors found that a Beltrami pseudosphere-shaped graphene sheet, for which all the (constant negative) curvature $\kappa$ is in the spatial part, leads in a straightforward way to analytic results. Their idea is that the electron quasiparticles close to one of the nondegenerate Dirac points of the first Brillouin zone of a Beltrami trumpet graphene sheet have local density of states (LDOS) conformally related to Rindler thermal-like LDOS. This is obtained theoretically by exploiting the Weyl scale symmetry in the case of graphene, an interesting topic by itself that has been studied in detail by Iorio \cite{i-2011}. The Rindler space is one that has a maximum acceleration and serves only to write down the power spectrum of the Rindler Green's function in the massless case, which is what a noninertial observer detects in this spacetime. The calculations involve the electronic Green functions, also known as Wightman functions in quantum field theories, whose analytic continuation just above the real axis are the LDOS up to the constant $-1/\pi$. In condensed matter physics the LDOS are directly  measurable quantities by scanning tunneling spectroscopy (STS). Iorio and Lambiase claim that the tip of the electronic microscope could detect the thermal-like Beltrami LDOS given by
$$
\rho_{B\kappa}(E,u) = \frac{|\kappa(u)|}{\pi^3}\frac{X_{\kappa}(E,u)}{e^{X_{\kappa}(E,u)}-1}
~,
$$
where $u$ is the spatial Beltrami coordinate, and the functions $\kappa(u)$ and $X_{\kappa}(E,u)$ are given as follows
$$
|\kappa(u)|=|\kappa|e^{-|\kappa|^{-\frac{1}{2}}u}~, \qquad X_{\kappa}(E,u)=\frac{1}{|\kappa|^{3/2}}\frac{2\pi E}{\hbar v_F}|\kappa(u)|~.
$$
This result can be also expressed in terms of an Unruh temperature function ${\cal T}(u)={\cal T}_0e^{|\kappa|^{1/2}u}$, where ${\cal T}_0=|\kappa|^{1/2}/2\pi$ can be determined from semilogarithmic plots of the experimental data. The thermal Bose-Einstein statistics for fermions is a well-known feature in odd spacetime dimensions.

\bigskip

On the other hand, Cveti\v c and Gibbons \cite{cg} put forward a (2+1)-dimensional gravity proposal for a graphene sheet in an external magnetic field, namely a Zermelo optical metric which is conformal to the BTZ black hole metric when the curvature of the sheet is negative. The BTZ black hole solutions in (2+1)-dimensional gravity have been discovered in 1992 by Ba\~nados, Teitelboim and Zanelli (BTZ) \cite{btz}. These solutions are characterized by two parameters, the mass $M$ and the angular momentum $J$, are locally anti-de Sitter and do not have any curvature singularity at the origin.
According to a general result, any stationary metric can be conformally rescaled into a Zermelo form
$$
ds^2=-dt^2 +h_{ij}(dx^i-W^idt)(dx^j-W^jdt),
$$
where $h_{ij}$ is the Zermelo optical metric and $W^i$ is the wind vector field\footnote{The terminology comes from the original minimax navigation problem studied by Zermelo in 1931 of finding the path between two points which minimizes the time traveled, given that one moves at unit speed with respect to the vector of the wind direction $W$.} that Cveti\v c and Gibbons identify with an induced gauge field due to an applied external magnetic field.
The Beltrami trumpet is the particular case corresponding to the naught wind vector and the simple polar optical metric $h_{ij}dx^idx^j=d\rho^2+C^2(\rho)d\phi^2$, $C(\rho)=a\exp(-\rho/a)$, $a$ - the constant radius of curvature, while the BTZ optical polar metric is slightly more complicated as it has a horizon geometry.

\bigskip

These remarkable proposals could have a rapid experimental progress due to the worldwide attention paid to graphene and other atomic-thin nanomaterials. However, even if these proposals look theoretically sound they can be more challenging to undertake in the lab than the theory would indicate. Based on recent nanoscience results scattered across the literature, we will describe in the following what has been really achieved in some laboratories on the lines suggested by these two proposals. We show that scientists wishing to turn these proposals into sound experiments confront a very complex and surprising experimental framework.

\bigskip

We want first to emphasize that some round features on the graphene sheets have been already produced in a number of laboratories.
In 2009, Stolyarova and collaborators reported the observation of graphene bubbles of diameters up to 30 nm and heights in the range of 0.5 - 2 nm on the surface of graphene flakes bombarded by a low-energy beam of protons \cite{stol}. Moreover, Georgiou {\em et al} \cite{geor} demonstrated that the shape of the bubbles can be controled by means of an external electric field, an experimental feat that could lead to real setups for the above-mentioned proposals. But we consider even more relevant, for both proposals, the scanning tunneling microscopy experiment conducted by
Levy {\em et al} \cite{levy} in the region of the bubbles formed in a graphene sheet grown on a platinum (111) surface. They detected peaks corresponding to Landau levels although no external magnetic field was applied. This revealed very strong pseudo-magnetic fields (up to 300 T) inside the bubbles which can be attributed to the induced vector potential generated by the strain applied to the graphene sheets. Thus, the unstrained common graphene Hamiltonian ($\sigma_i$ are the Pauli matrices)
$$
H_0= v_F(\sigma_1 p_1+\sigma_2 p_2)
$$
is changed to a strained Hamiltonian
$$
H_S=v_F(\vec{p}-e\vec{A}_S)\cdot \vec{\sigma}~,
$$
where the main contribution to the strain-induced vector potential comes mainly from local perturbations to the hopping amplitudes between the $\pi$ orbitals in nearest-neighbor atoms. From the observed magnetic lengths in the range $l_B=1.5-2$ nm the pseudomagnetic fields $B(T)=26^2l_{B}^{-2}$ (nm) are in the range 100-300 $T$.
Thus, taking into account that  Cveti\v c and Gibbons suggest to apply a real external magnetic field in order to induce a Zermelo metric, the interaction with the pseudomagnetic field created by the strain-induced BTZ graphene sheet should be taken into account in the calculations. The interplay between the two fields is currently a hot topic in graphene physics \cite{inter}. As a matter of fact, because of the pseudomagnetic fields, the conditions to achieve zero Zermelo wind vector fields could be obtained in principle at nonzero applied magnetic fields.

\bigskip

As commented by Iorio and Lambiase, some experimental conditions should be imposed for the observation of the analogue effects. The most important one requires that the measuring device, i.e., the tip of the cantilever, should very closely follow the profile of the graphene trumpet in order to move as much as possible as a noninertial detector. From the radiometric standpoint, this implies that we are in the {\em near-field regime} since the measurements are performed at distances smaller than the thermal
wavelength $\lambda_{th}= \hbar c/k_BT$ \cite{yau}, which at temperatures less than the room temperature is bigger than 7.6 $\mu$m.
It is well established since more than forty years by now \cite{1971} that in this regime there is only a small contribution of the propagating modes to the radiative heat flux which is dominated by evanescent waves, especially by surface polaritons (mixtures of a photon with an excitation of the material such as a phonon or a plasmon)\cite{2007}. The latter can produce resonant energy transfer restricted to small frequency bands around the surface mode resonance frequency \cite{vafek,stauber}.
Moreover, there is already strong evidence that the graphene electron quasiparticles can couple to plasmons forming the so-called plasmarons. Very recently, Carbotte {\em et al} described their behavior in near field optics \cite{carb}.

\bigskip

On the other hand, both Hawking and Unruh effects are far-field thermal-like effects, i.e., only propagating modes are taken into account and maximum transmission for all allowed frequencies is assumed. However, in the case of nanoscale experiments, roughness can couple surface waves to propagating waves transferring the coherence properties of collective surface waves into the far field. The intensity statistics is still very different from that of a coherent beam but nevertheless the surface waves can produce a significant time and spatial second order coherence of the field in the near-field.

\bigskip

Since radiative heat experiments in the case of graphene are naturally near-field measurements involving its collective surface excitations, we believe that one can explore at the minute nanometer scales processes predicted by the more electromagnetic-like membrane paradigm rather than Hawking-Unruh paradigm. Clues in this sense can be found in the similarity of the  terminology. For example, it is worth noticing that Wakker {\em et al} \cite{wtb} have suggested that pseudo-magnetic fields may result in circulating probability currents whereas the eddy currents have been one of the original ideas in the first papers on the membrane paradigm by Damour \cite{d78} and Znajek \cite{z78}.

\bigskip

To conclude, the rapid advance in strain engineering of graphene can be used not only as a promising alternative tool in graphene electronics but also to disentangle some of the most cherished predictions of physics in extreme gravitational and high energy conditions. Since in the case of graphene experiments we are essentially in near field conditions, they could reveal analogues of phenomena more similar to those predicted by the membrane paradigm rather than the asymptotic thermal-like effects of Hawking and Unruh that belongs to the far-field conditions.

\bigskip

\noindent {\small Acknowledgment}: PC appreciates the supports by Taiwan National Science Council(NSC) under Project No. NSC-100-2119-M-002-525, No. NSC-100-2112-M-182-001-MY3, US Department of Energy under Contract No. DE-AC03-76SF00515, and the NTU Leung Center for Cosmology and Particle Astrophysics (LeCosPA). 
HR thanks CONACyT-Mexico for a sabbatical fellowship and NTU LeCosPA for the kind hospitality.


\begin{thebibliography}{123}

\bibitem{livrev} C. Barcelo, S. Liberati, M. Visser, {\em Analogue Relativity}, Living Rev. Rel. 8, 12 (2005).
\bibitem{il-2011} A. Iorio, G. Lambiase, {\em The Hawking-Unruh phenomenon on graphene}, Phys. Lett. B 716, 334 (2012). 
\bibitem{i-2011} A. Iorio, {\em Weyl-gauge symmetry of graphene}, Ann. Phys. 326, 1334 (2011).
\bibitem{cg} M. Cveti\v c, G.W. Gibbons, {\em Graphene and the Zermelo optical metric of the BTZ black hole}, Ann. Phys. 327, 2617 (2012). 
\bibitem{btz} M. Ba\~nados, C. Teitelboim, J. Zanelli, {\em Black hole in three-dimensional spacetime}, Phys. Rev. Lett. 69, 1849 (1992).
\bibitem{stol} E. Stolyarova et al., {\em Observation of graphene bubbles and effective mass transport under graphene films}, Nano Lett. 9, 332 (2009).
\bibitem{geor} T. Georgiou et al., {\em Graphene bubbles with controllable curvature}, Appl. Phys. Lett. 99, 093103 (2011).
\bibitem{levy} N. Levy et al., {\em Strain-induced pseudo-magnetic fields greater than 300 Tesla in graphene nanobubbles}, Science 329, 544 (2010).
\bibitem{inter} K.-J. Kim, Ya. M. Blanter, K.-H. Ahn, {\em Interplay between real and pseudomagnetic field in graphene with strain},
Phys. Rev. B 84, 081401 (2011).
\bibitem{yau} Z. Yan, {\em General thermal wavelength and its applications}, Eur. J. Phys. 21, 625 (2000).
\bibitem{1971} D. Polder, M. Van Hove, {\em Theory of radiative heat transfer between closely spaced bodies}, Phys. Rev. B 4, 3303 (1971).
\bibitem{2007} A.I. Volokitin, B.N.J. Persson, {\em Near-field radiative heat transfer and noncontact friction}, Rev. Mod. Phys. 79, 1291 (2007).
\bibitem{vafek} O. Vafek, {\em Thermoplasma polariton within scaling theory of single-layer graphene}, Phys. Rev. Lett. 97, 266406 (2006).
\bibitem{stauber} T. Stauber, G. G\'omez-Santos, {\em Plasmons and near-field amplification in double-layer graphene}, Phys. Rev. B 85, 075410 (2012).
\bibitem{carb} J.P. Carbotte, J.P.F. LeBlanc, E.J. Nicol, {\em  Emergence of plasmaronic structure in the near field optical response of graphene},
Phys. Rev. B 85, 201411(R) (2012). 
\bibitem{wtb} G.M.M. Wakker, R.P. Tiwari, M. Blaauboer, {\em Localization and circulating currents in curved graphene devices}, Phys. Rev. B 84, 195427 (2011).
\bibitem{d78} T. Damour, {\em Black hole eddy currents}, Phys. Rev. D 18, 3598 (1978).
\bibitem{z78}    R.L. Znajek, {\em The electric and magnetic conductivity of a Kerr hole}, Mon. Not. R. astr. Soc. 185, 833 (1978).

\end{thebibliography}
\end{document}